\documentstyle[aps, preprint]{revtex}
\tightenlines

\def\be{\begin{equation}}
\def\ee{\end{equation}}
\def\ba{\begin{eqnarray}}
\def\ea{\end{eqnarray}}
\def\half{{1\over 2}}
\def\x{\mbox{\bf x}}
\def\k{\mbox{\bf k}}
\def\p{\mbox{\bf p}}
\def\R{\mbox{\bf R}}

\def\aPhi{\langle\Phi\rangle}
\def\aPsi{\langle\Psi\rangle}
\def\aphi{\phi_0}
\def\Hub{{\cal H}}
\def\Z{{\cal Z}}

\input epsf
\draft
\begin{document}
\preprint{PURD-TH-01-02, hep-ph/0107143}
\title{Metric perturbations at reheating: the use of spherical symmetry}
\author{F. Finelli and S. Khlebnikov}
\address{Department of Physics, Purdue University,
West Lafayette, IN 47907, USA}
\date{July 2001}
\maketitle
\begin{abstract}
We consider decay of the inflaton with a quartic potential coupled to
other fields, including gravity, but restricted to spherical symmetry.
We describe analytically an early, quasilinear regime, during which inflaton 
fluctuations and the metric functions are driven by nonlinear effects of
the decay products. We present a detailed study of the leading nonlinear
effects in this regime. Results of the quasilinear approximation, in its
domain of applicability, are found to be consistent with those of fully 
nonlinear lattice studies. We discuss how these results may be promoted
to the full three dimensions.
\end{abstract}

\section{Introduction}
Reheating is a crucial ingredient of inflationary cosmology \cite{BOOKS}.
In most models of inflation, the quasi-exponential expansion leaves
the universe very cold and empty, with energy stored in coherent 
oscillations of the inflaton field. Some kind of nonlinear process is 
required to convert that energy into energy of ordinary particles.

Products of the inflaton decay (which will be denoted as
$\chi$ but in general will include the short-wave fluctuations of the 
inflaton field itself) can cause metric perturbations in a variety of ways.
Two possible processes are represented by diagrams of Fig. \ref{fig:diag}.
In the first process, $\chi$ fluctuations scatter off the inflaton's
homogeneous component (which is still large at that time), thereby
producing inflaton fluctuations, which in turn cause perturbations of the
metric. In the second process, $\chi$ fluctuations drive the metric 
perturbations directly.

We want to stress that the processes shown in the diagrams are quite generic
and exist regardless of the particular mechanism of reheating. 
They exist even if the inflaton only decays
perturbatively, although the scattering process of Fig. 1a is absent from 
the early-days perturbative
treatment \cite{early}, in which the decay is represented by an effective 
friction term in the equation of motion for the inflaton field $\phi$. 
Fig. 1a corresponds to 
a driving force in the equation of motion for $\phi$ that 
may allow fluctuations of $\phi$ to reach values $\delta\phi \sim \phi$.
The question remains of course (which we will address in this paper)
what portion of this large fluctuation resides 
in super-Hubble modes, in particular those relevant to structure formation.

Recent progress in the theory of reheating came after recognition that
in many inflationary models the early stages of reheating can be
studied rather completely. These are models where reheating starts by 
a parametric resonance \cite{KLS,STB}. Parametrically amplified fluctuations 
are almost classical, and nonlinear effects can be explored numerically in the
classical approximation \cite{KT1}. It was found that in many cases
the resonant stage is followed by a stage (called semiclassical thermalization),
during which nonlinear interactions of fluctuations become increasingly important, 
and power spectra of the fields eventually become smooth \cite{KT1,KTPLB,KT2}.
Importance of the scattering process 
of Fig. 1a was recognized early on in lattice studies in {\em rigid spacetimes}
\cite{KT1}, and since then the
estimate $\delta\phi \sim \mbox{(0.1--1)}\phi$ has been confirmed for a number 
of models \cite{KT2,KTGRAV,PR,LECTURES}.

Meanwhile, applications of the linear theory of metric perturbations to
a special type of models have shown that in principle also super-Hubble
modes, which necessarily activate {\em gravity}, can undergo growth
at reheating \cite{BKM,FB1,FB2,BV}. 
It is therefore became important to combine the 
two lines of research and see if a similar super-Hubble growth can occur due to 
nonlinear effects. The first attempts is that direction were made by
studies in restricted geometries: planar geometry in Ref. \cite{EP} and
spherical symmetry in Ref. \cite{FK}. Both studies have indeed found
such a growth \cite{EP,FK}.

The main purpose of the present paper is to supply an analytical
counterpart to the numerical method of Ref. \cite{FK}. Although our 
approximate method, which we call {\em quasilinear} theory, cannot describe
the fully nonlinear stage of the evolution, it allows us to describe
the relevant nonlinear effects at earlier times, when the nonlinearity is still
weak. This method retains linear terms in fluctuations of $\phi$ and the 
metric and quadratic terms in $\chi$. It generalizes the method developed in
Refs. \cite{KT2,LECTURES} for rigid spacetimes to the case when metric perturbations 
are present.
In particular, it will allow us to take into account the processes shown in
Fig. 1, which are not included in the standard linear theory of metric
perturbations. (In this paper, we consider the case when the expectation value
of $\chi$ is initially zero, so these nonlinear effects are the leading effects
of $\chi$.)

The system of spherically-symmetric equations is written down in
Sect. \ref{sect:sphe}. In Sect. \ref{sect:quasi}, we construct the quasilinear 
theory, restricting our
attention to a model with a quartic inflaton potential. Within the quasilinear
theory, we identify two potential sources of metric perturbations at super-Hubble
scales. The first is the ``erosion'' of the inflaton by $\chi$
fluctuations, as shown in Fig. 1a; the second, is the metric perturbations
caused by the anisotropic stress, a purely relativistic effect that can be viewed
as a part of Fig. 1b. In the subsequent sections
we analyze these effects in turn. We find that unless the decay products themselves
have infrared-singular spectra the ``erosion'' is not
infrared singular and thus cannot be important on scales relevant to
structure formation. On the other hand, the anisotropic stress gives rise
to long-range (secular) terms in the gravitational potentials and is
important at large scales, in a calculation
restricted to spherical symmetry. In the concluding Sect. \ref{sect:conc}, 
we discuss the
overall physical picture that emerges when we try to promote our findings
to the full three dimensions.

\section{Gravitating scalar fields in spherical symmetry}
\label{sect:sphe}

Early on in reheating, the particle density is small, and a weakly interacting
particle species has a very large mean-free path. For such a species, 
a spherically-symmetric field will be 
relatively stable with respect to fragmentation into higher partial
waves. This means that the restriction to spherical symmetry, when it is
desired or unavoidable, can be imposed directly on the fields (and the
metric), rather than on compound quantities, such as the stress tensor.
In this sense, our calculations will be quite different from those with
spherically-symmetric {\em fluids}: in the latter case particles move in all
directions; it is only the large-scale motion of the fluid that is radial.
In our case, because of the large mean-free paths, the fluid approximation
is not readily available.

We consider a system of scalar fields minimally coupled to gravity and 
described by the Lagrangian
\be
S = \int d^4 x \sqrt{-g} \left[ \frac{R}{2 \kappa^2} - \frac{1}{2}
\sum_i g_{\mu \nu} \partial^\mu \phi_i \partial^\nu \phi_i - V \right]
\ee 
where $i$ labels different fields, $\kappa^2 = 8\pi G$, 
and $V = V(\phi_i) $ is the potential which can depend on all of them.
The most general spherically symmetric metric is 

\be
ds^2 = (- N^2 + \psi^4 \beta^2) dt^2 + 2 \psi^4 \beta dt dr + \psi^4 dr^2 + 
\alpha^2 r^2 d\Omega^2
\ee
where $d\Omega^2$ is the line element of the 2-sphere and all the four 
functions $N, \psi, \alpha, \beta$ depend on $t$ and $r$. 
We shall work in {\em isotropic} coordinates \cite{choptuik} ($\alpha^2 =
\psi^4$). 
Accordingly we shall have only one conjugate variable to $\psi$ by    
a diagonal gauge in which the shift vector $\beta$ is set to zero: in 
this way the extrinsic curvature will be determined only by its trace
\cite{choptuik}. 
Therefore the metric we shall use is the following

\be
ds^2 = - N^2(t, r) dt^2 + \psi^4 (t, r) \left[ dr^2 + r^2 d \Omega^2
\right] \,.
\label{metric}
\ee
The equation of motion for a scalar field with a generic potential $V(\phi)$ 

\be 
- \nabla^\mu \partial_\mu \phi_i + \frac{\partial V}{\partial \phi_i} = 0
\ee
are written in a Hamiltonian form for the metric (\ref{metric})
\begin{eqnarray}
\dot \phi_i &=& N \frac{\pi_i}{\psi^6} 
\label{dotphi}
\\
\dot \pi_i &=& \frac{1}{r^2}\left( r^2 \psi^2 N \phi_i' \right)' - N
\psi^6 \frac{\partial V}{\partial \phi_i}
\label{dotpi}
\end{eqnarray}
where $\phi, \pi_i$ are the pair of conjugate variables for the 
scalar field, a dot denotes the derivative with respect to $t$ whereas 
a prime denotes the derivative with respect to $r$.

The Hamiltonian equations for the metric are the following:

\begin{eqnarray}
\dot \psi &=& - \frac{N \psi}{6} K \\
\frac{\dot K}{N} &=& \frac{K^2}{2} -6 \frac{\psi'}{\psi^5}
\left( \frac{\psi'}{\psi} + \frac{1}{r} \right) - 3 \frac{N'}{N \psi^4}
\left( \frac{1}{r} +
2 \frac{\psi'}{\psi} \right) + \frac{3}{2} \kappa^2 E - 3 \kappa^2 V
\label{Kequation}
\end{eqnarray}
where $K$ is the trace of the extrinsic curvature and
$E$ is the total Hamiltonian for the scalar fields
\be
E = \sum_i \left(
\frac{\pi_i^2}{2 \psi^{12}} + \frac{\phi_i\,'^2}{2 \psi^4} \right) +
V
\,.
\label{energyphi}
\ee
The equation for the lapse function $N$ does not involve time derivatives
(since $N$ is not a dynamical degree of freedom):
\be
\frac{r \psi^4}{N} \left( \frac{N\,'}{r \psi^4} \right)\,' - 6
\frac{\psi\,'^2}{\psi^2} + \frac{2 r}{\psi} \left( \frac{\psi\,'}{r}
\right) \,' = - \kappa^2 \sum_i \phi_i\,'^2 
\label{Nequation}
\ee 

The energy and momentum constraints are
\begin{eqnarray}
\frac{1}{2} \, {^{(3)} R} = - \frac{4}{\psi^5} \, \nabla^2 \psi = \left[
\kappa^2 E - \frac{K^2}{3} \right] \\
\label{enconstraint}
\frac{2}{3} K' = \sum_i \kappa^2 \frac{\pi_i \phi_i\,'}{\psi^6} 
\label{momconstraint}
\end{eqnarray}
where ${^{(3)} R}$ is the curvature of 
spatial sections.

\section{Quasilinear regime} \label{sect:quasi}

To describe analytically the phase that follows the linear 
stage, we employ now a {\em quasilinear} approximation, which generalizes
the method developed in \cite{KT2,LECTURES} for rigid spacetimes.
In this approximation we consider terms that are linear in
the fluctuations of the inflaton and of the metric, 
but keep quadratic terms in $\chi$ fluctuations. In the absence of an
initial homogeneous value of $\chi$, these quadratic terms represent the
leading effects of $\chi$ on the inflaton and the metric functions.

The metric at the first order has the same general form as in the linear
theory (reviewed in Ref. \cite{MFB}). Including only scalar perturbations 
but keeping (for now) three-dimensional coordinates, we write it in the 
longitudinal gauge as
\be
ds^2 = a^2(t) \left\{ -[1-2 \Phi(t, {\bf x})] dt^2 + [1 + 2 \Psi
(t, {\bf x})] d{\bf x}^2
\right\} \, ;
\label{quasimetric}
\ee$t$ now is the conformal time.
An important difference with the linear theory, however, is that even
for minimally coupled scalars, there is an {\em
anisotropic stress}, induced by a nonlinear contribution of $\chi$.
As a result,  $\Phi \neq \Psi$.

For definiteness, from now on we consider a model with two fields: 
the inflaton
$\phi$ and the second field $\chi$ whose fluctuations will 
drive fluctuations of $\phi$ and the metric.
Both fields are minimally coupled to gravity, and their potential is
\be
V(\phi,\chi) = {1\over 4} \lambda\phi^4 + \half g^2 \phi^2\chi^2 
+ {1\over 4} \lambda'\chi^4 \; .
\label{pot}
\ee
Our method applies equally well to single field models, in which
case the role of $\chi$ will be played by short-scale components of the
inflaton itself.

To the linear order in metric perturbations, equations for the fields
$\phi$ and $\chi$ 
can be obtained without restriction to spherical symmetry.
On the other hand, equations for the metric functions we present below 
are valid only under that restriction.

Expanding the equation of motion for $\phi$ to the first order in
$\Phi$, $\Psi$, $\chi^2$, and the spatial derivatives of $\phi$ we obtain
\be
\ddot \phi + 2 \Hub \dot \phi +  \nabla^2 \phi + \lambda a^2 \phi^3 
+ g^2 a^2 \phi \chi^2 = 
- \dot \phi (\dot \Phi + 3 \dot \Psi) + 2 \lambda a^2 \phi^3 \Phi \; ,
\label{fophi}
\ee
where $\Hub = \dot a / a$; dots denote derivatives with respect to $t$.
We can separate the field $\phi$ into
its volume average, $\phi_0$, and a fluctuation:
\be
\phi(t,\x) = \phi_0(t) + \delta \phi(t,\x) \; .
\label{dphi}
\ee
The volume averages of $\Phi$ and $\Psi$ are made zero through
definitions of the scale factor $a$ and the conformal time. We also assume 
that the volume average of the field $\chi$ is initially zero.

Substituting (\ref{dphi}) into (\ref{fophi}), expanding to the first
order, and averaging over the volume gives
\be
\ddot \phi_0 + 2 \Hub \dot \phi_0 + \lambda a^2 \aphi^3 
+ g^2 a^2 \aphi \overline{\chi^2} = 0 \; ;
\label{quasiave}
\ee
the overline denotes a volume average.
Subtracting (\ref{quasiave}) from (\ref{fophi}), we obtain an equation for
the fluctuation:
\be
\delta \ddot \phi + 2 \Hub \delta \dot \phi + \left[ -\nabla^2
+ 3 \lambda a^2 \aphi^2 \right] \delta\phi 
+ g^2 a^2 \aphi  (\chi^2 - \overline{\chi^2}) =
-  \dot \phi_0 (\dot \Phi + 3 \dot \Psi) + 2\lambda a^2 \aphi^3 \Phi \; .
\label{quasiphi}
\ee
Similarly, for $\chi$, to the first order in $\Phi$, $\Psi$, $\chi^2$,
and $\delta\phi$, we obtain
\be
\ddot \chi + 2 \Hub \dot \chi +  \left[-\nabla^2  + g^2 a^2 \aphi^2 \right] \chi =
- 2  g^2 a^2 \aphi\chi\delta\phi  - \dot \chi (\dot \Phi + 3 \dot \Psi) \; .
\label{fochi}
\ee
In this paper, we use quasilinear theory only for the earliest times,
when the right-hand of (\ref{fochi}) can be neglected. In this case, we
obtain an equation for $\chi$ which is unaffected by metric perturbations:
\be
\ddot \chi + 2 \Hub \dot \chi +  \left[-\nabla^2 + g^2 a^2 \aphi^2 \right]\chi 
= 0 \; .
\label{quasichi}
\ee

In the full 3d case, the nonlinearities of $\chi$ would
act as a source not only for scalar metric perturbations, but also for vector
and tensor ones. Equations given below are restricted to
spherical symmetry, in which case only scalar perturbations are 
present.

In the quasilinear approximation, 
the energy constraint (\ref{enconstraint}) reads
\be
- \nabla^2 \Psi + 3 \Hub (\dot \Psi + \Hub \Phi) = 
\frac{\kappa^2}{2} \left( \delta
E_{\phi} + \delta E_{\chi} \right) 
\ee
where the fluctuations of energy densities of the fields
are defined by $\delta E_i = E_i - \overline{E_i}$; for the model at hand, to
the required order,
\begin{eqnarray}
\delta E_{\phi} & = & \dot \phi_0 \delta \dot{ \phi}
+ \dot \phi_0^2 \Phi + a^2 \lambda \phi_0^3 \delta \phi \; , \\
\label{ephi}
E_{\chi} & = & \frac{\dot{ \chi}^2}{2} + \frac{(\nabla
\chi)^2}{2} + \frac{g^2}{2} a^2 \phi_0^2  \chi^2
\label{echi} \; .
\end{eqnarray}
The momentum constraint (\ref{momconstraint}) reads
\be
( \dot \Psi + \Hub \Phi )\,' = - \frac{\kappa^2}{2} ( \dot \phi_0 \delta
\phi\,' + \dot \chi  \chi\,' ) \; .
\label{quasimc}
\ee
We do not present a quasilinear form for Eq.
(\ref{Kequation}). The quasilinear form for Eq. (\ref{Nequation}) is:
\be
r \left( \frac{(\Phi - \Psi)'}{r} \right)' = \kappa^2 {\chi\,'}^{2} \, .
\label{quasiN}
\ee
The right-hand side is the anisotropic stress.

In general, an {\em anisotropic stress} is generated by a non-vanishing 
traceless part of the spatial energy-momentum tensor. In our case, it is
due to the traceless second-order contribution of the field $\chi$:
\be
\Pi^i_j = \partial^i \chi \partial_j \chi - \frac{1}{3}
\delta^i_j \partial^l \chi \partial_l \chi \; .
\ee
For small scalar perturbations, the corresponding Einstein equation takes
\be
( D^i D_j - \frac{1}{3} \delta^i_j D^2 ) \left[ \Phi - \Psi \right] =
\kappa^2 \left[ \Pi^i_j \right]_S \, ,
\label{scalar}
\ee
where $D_i$ is the covariant derivative with respect to the spatial metric
$\gamma_{ij}$, related to $g_{ij}$ via $g_{ij}=\psi^4 \gamma_{ij}$,
$D^2 = D^i D_i$, and the subscript $S$ denotes the scalar part of $\Pi^i_j$. 
The scalar part is automatically selected when we make restriction to
spherical symmetry, and the result coincides with Eq. (\ref{quasiN}).
 
We see that nonlinearities in $\chi$ can potentially 
cause two types of effects. First, the rescattering term---the 
last term on the right-hand side of (\ref{quasiphi})---will lead to a growth
of $\delta\phi$; this is the effect represented by the diagram of Fig. 1.
Second, the right-hand side of Eq. (\ref{quasiN}) directly produces
$\Phi-\Psi \neq 0$. Either way, the resulting perturbations are non-Gaussian.
In the remaining sections, we will study these two effects in turn.
Note that {\em neither} of these effects was taken into account in 
Ref. \cite{LLMW} where only the direct second-order contribution of $\chi$ 
to curvature perturbation was considered (in a different model).

\section{Inflaton fluctuations} \label{sect:infl}

The two effects described at the end of the previous section are more easily
disentangled if
we use, instead of $\delta\phi$, the gauge-invariant variable
\be
Q_\phi = \delta \phi - \dot \phi \Psi/\Hub \; .
\label{Qphi}
\ee
This variable was introduced by Mukhanov to characterize perturbations produced 
{\em during} inflation \cite{MUKHA85}; it was applied to studies of reheating
at the linear level in Refs. \cite{KHNT,FB1,FB2,BV}.\footnote{This variable
looks as a combination of metric 
and field fluctuations in the longitudinal gauge, 
while in the uniform curvature
gauge its interpretation is more transparent: it is simply
the field fluctuation, which indeed is the true dynamical variable of the
problem (see for example \cite{hwang}).}
From now on we omit the zero subscript on the homogeneous component of $\phi$,
hoping that this will not cause any confusion. 

Using Eq. (\ref{quasiphi}) and the quasilinear equations for the metric, 
we obtain the following equation for $Q_\phi$:
\be
\ddot Q_\phi + 2 \Hub \dot Q_\phi + \left[ -\nabla^2 + 3 \lambda a^2
\phi^2 + \omega^2_{\rm GRAV} \right] Q_\phi 
+ g^2 a^2 \phi \delta\chi^2 = F (\chi) \,, 
\label{qphi}
\ee
where $\delta\chi^2 = \chi^2 - \overline{\chi^2}$, 
the gravitational contribution $\omega^2_{\rm GRAV}$ to the
frequency is
\be
\omega^2_{\rm GRAV} = \kappa^2 \dot \phi \left( 2 a^2 \frac{V_{,\phi}}{\Hub} 
+ 3\dot \phi - \kappa^2 \frac{\dot \phi^3}{2 \Hub^2} \right) \,, 
\label{modgrav}
\ee
and $V$ is the potential of $\phi$; in our case, 
$V_{,\phi} = \lambda\phi^3$. This potential does not include a Hartree 
correction, cf. Eq. (\ref{quasiave}). Indeed, including such a correction here
would take us outside of the quasilinear approximation. Thus, (\ref{modgrav})
is the standard expression of the linear theory \cite{MUKHA85,MFB,FB1}.
The force term $F (\chi)$ depends only on $\chi$:
\begin{eqnarray}
F(\chi) & = & 
  \kappa^2 \frac{\dot \phi}{ \Hub} \left[ - \frac{1}{3} (\nabla\chi)^2 
  - {1\over 2} g^2 a^2 \phi^2 \delta\chi^2 
  + \frac{1}{3\kappa^2} \nabla^2 (\Phi - \Psi) \right]  \nonumber \\
& & 
  - \left( 6\dot{\phi} + 2a^2\frac{V_{,\phi}}{\Hub} 
  - \kappa^2 \frac{\dot{\phi}^3}{\Hub^2} \right) \Z_\chi \; ,
\label{forcegrav} 
\end{eqnarray}
where
$\Z_\chi$ is the contribution of $\chi$ to the curvature perturbation,
see eqs. (\ref{mc3d}), (\ref{zetaquasi}) below. The term proportional 
to $\nabla^2 (\Phi - \Psi)$
depends only on $\chi$ by virtue of the quasilinear approximation, 
see Eq. (\ref{quasiN}).

Because Eq. (\ref{qphi}), unlike Eq. (\ref{quasiphi}), depends on the gravitational 
potentials only through $\nabla^2 (\Phi - \Psi)$, the most long-range (secular)
contributions to $\Phi$ and $\Psi$ do not enter, and all interactions generated by
gravity in Eq. (\ref{qphi}) are suppressed by at least one power of $M_{\rm Pl}$.
(This suppression is in parallel with the result of Ref. \cite{FB2} for the linear
case.)
Neglecting all such terms and going over to the variable $X=a Q_\phi$, we obtain
\be
\ddot X + (-\nabla^2 + 3 \lambda a^2 \phi^2) X
= - g^2 a^3 \phi (\chi^2 - \overline{\chi^2}) \; ,
\label{quasiX}
\ee
where  we have neglected the term $\ddot a/a$ in the frequency because a
radiation dominated universe ($\ddot a \approx 0$) is a good approximation
during reheating in this model.
Eq. (\ref{quasiX}) shows that, in the quasilinear theory, the growth of
$X$ is driven primarily by rescattering due to the local coupling and is 
essentially unaffected by gravity.

By using the spherically symmetric equations for gravity in deriving 
Eq. (\ref{quasiX}), we have neglected vector and tensor perturbations.
It seems plausible, however, that contributions of these types of perturbations
to the evolution of $Q_{\phi}$ are suppressed in the same way as the 
contribution
of scalar perturbations. We therefore expect that Eq. (\ref{quasiX}) applies,
to the leading order in $M_{\rm Pl}^{-1}$, in the full three dimensions.
Accordingly, we write it in the Fourier space as
\be
\ddot X_{\bf k} + (k^2 + 3 \lambda a^2 \phi^2) X_{\bf k} 
= - g^2 a^3 \phi \int d^3 p \chi_{\bf p}(t') \chi_{\bf k+p}(t') \; .
\label{forced}
\ee
The solution to Eq. (\ref{forced}) can be written in terms of
the Green function:
\be
X_{\bf k} (t) = X_{{\bf k}, \rm INFL} (t)
- g^2 \int_{0}^{t} dt' G_k (t,t') a^3 (t') \phi (t')
\int d^3 p \chi^*_{\bf p}(t') \chi_{\bf k+p}(t') \, ,
\label{sol}
\ee
where $X_{{\bf k}, \rm INFL} (t)$ is the solution to the linear problem,
corresponding to perturbations produced during inflation and the crossover
to reheating. The second, forced,
term in Eq. (\ref{sol}) starts from zero at $t=0$ but then grows, initially
as $\exp({2 \mu_{\max} t})$ where
$\mu_{\max}$ is the maximal Floquet exponent for $\chi$. There is no 
additional decay in time, since the factor $a^3$ cancels the damping
envelope $\sim 1/a$ of the fields $\phi, \chi$.

The Green function can be written as
\be 
G_k (t,t') = \frac{X_{k\, 1}^{(0)} (t') X_{k\, 2}^{(0)} (t) 
- X_{k\, 1}^{(0)} (t) X_{k\, 2}^{(0)} (t')}{W_k} \,,
\label{green}
\ee
where $X_{k\, 1,2}^{(0)}$ are two independent solutions to the
homogeneous equation
\be
\ddot X_{k \, 1,2}^{(0)} (t) 
+ (k^2 + 3 \lambda a^2 \phi^2) X_{k \, 1,2}^{(0)} = 0 \; ,
\ee
and $W_k$ is their (time independent)
Wronskian. The homogeneous solutions are known in implicit form
\cite{GKLS} (see also \cite{kaiser}):
\be
X_{{\bf k} \, 1,2}^{(0)} (t) = \sqrt{|M_k (t)|} 
\exp \left( \pm \frac{W_k}{2} \int \frac{dt}{M_k(t)} \right) \; ,
\label{solution}
\ee
where $M_k(t)$ is defined in Eq. (37) of Ref. \cite{GKLS}, and 
the Wronskian $W_k$ is
\be
W_k = \frac{8}{9} k \sqrt{(x^4 - \frac{9}{4})(3 - x^4)} 
\ee
with $x=k/(\sqrt{\lambda} \phi(0))$.

The Wronskian $W_k$ is real in the single
resonance band $3/2 < x^2 < \sqrt{3}$ and 
imaginary everywhere else. When $W_k$ is imaginary, $W_k=i|W_k|$,
the solutions (\ref{solution}) are
complex conjugate, and the Green function becomes
\be
G_k (t,t') = {2\over |W_k|} \sqrt{|M_k (t) M_k(t')|} 
\sin\left( \frac{|W_k|}{2} \int_t^{t'} \frac{dt''}{M_k(t'')} \right) \; .
\label{green1}
\ee
At small $k$, $|W_k| \sim k$, while $M_k(t)$ is finite and nonzero, so
$G_k$ is regular there. Thus,
if the field $\chi$ has not been amplified during inflation, the integral
over ${\bf p}$ in Eq. (\ref{sol}) remains regular at $k=0$ as long as 
the quasilinear approximation applies. 

Using Eq. (\ref{sol}), we can estimate the power spectrum of the
forced (non-Gaussian) part of $Q_{\phi}$ as
\be
P_{Q,{\rm NG}}(k) \sim \left( \frac{g^2}{\lambda\phi} \right)^2
\int d^3 p P_{\chi}(p) P_{\chi}(|\p + \k|) \; ,
\label{pwsQ}
\ee
where $P_{\chi}$ is the power spectrum of $\chi$. We should now distinguish 
two cases.

(i) If the field $\chi$ was light during inflation, $g^2 \alt \lambda$, the
spectrum of $\chi$ produced at that time is infrared singular, which by
Eq. (\ref{pwsQ}) gives rise to an infrared-singular spectrum of $Q_{\phi}$.
This will result in non-Gaussian adiabatic perturbations of the metric at
scales of relevance to structure formation, in addition to the previously
considered \cite{LM} perturbations due to a direct quadratic
contribution of $\chi$ to the energy density.

For $1 < g^2 /\lambda < 3$, when the infrared part of the spectrum is 
amplified by parametric resonance, this effect can be quite large. It is the
largest for $g^2 /\lambda = 2$, when the resonance peak is centered at $k=0$.
In this case, assuming for simplicity the exact scale-invariant spectra
for both $\phi$ and $\chi$ at the end of inflation, we can estimate
$P_{\chi}$ as $P_{\chi}(k) \sim \langle\chi^2 \rangle_{\rm full} / k^3$,
where $\langle\chi^2 \rangle_{\rm full}$ denotes the full 3d variance.
The resulting $P_{Q,{\rm NG}}$ should be compared to the 
inflation-produced Gaussian spectrum $P_{Q,{\rm INFL}}\sim H^2_{\rm INFL} / k^3$, 
where $H_{\rm INFL}$ is the Hubble parameter during inflation. Using
the maximal value of the variance 
$\langle\chi^2 \rangle_{\rm full} \sim \lambda \phi^2 / g^2$ \cite{KT2,KTGRAV}, 
we obtain
\be
\frac{P_{Q,{\rm NG}}}{P_{Q,{\rm INFL}}} 
\sim \left( \frac{g^2}{\lambda\phi} \right)^2
\frac{\langle\chi^2 \rangle_{\rm full}^2}{H^2_{\rm INFL}}
\sim \frac{\phi^2}{H^2_{\rm INFL}} \sim 10^{6} \; .
\ee
(A similar, but Gaussian, effect is obtained in linear theory in certain 
models with a nonvanishing homogeneous component of $\chi$ \cite{FB2,BV}.)

(ii) If the field $\chi$ was heavy during inflation, it arrives at reheating 
with a constant (white-noise) spectrum at small $k$, and by Eq. (\ref{pwsQ}) 
the spectrum of $Q_{\phi}$ is also constant (``non-Gaussian white noise''). 
Indeed, assuming that the resonance peak is located at $k=k_*$, we estimate that
$P_{\chi} \sim \langle \chi^2 \rangle_{\rm full} / k_*^3$, where 
$\langle \chi^2 \rangle_{\rm full}$ denotes the full 3d variance. Then,
\be
P_{Q,{\rm NG}} \sim \left( \frac{g^2}{\lambda \phi} \right)^2 
\frac{\langle \chi^2 \rangle_{\rm full}^2}{k_*^3} \, .
\label{ngestimate}
\ee  
Using again the maximal value
$\langle \chi^2 \rangle_{\rm full} \sim \lambda \phi^2 / g^2$, we obtain
\be
P_{Q,{\rm NG}} \sim \phi^2 / k_*^3 \; .
\label{pwsQ2}
\ee

This estimate can be converted into an estimate for curvature perturbations,
as measured by the parameter $\zeta$,
\be
(1+w) \zeta = \frac{2}{3} \left( \dot{\Psi} / \Hub + \Phi \right) 
+ (1+w)\Psi \; ,
\label{zetadef}
\ee
where
\be
1+w = \frac{\kappa^2 {\dot \phi}^2}{3 \Hub^2} \; .
\ee
The term $1 + w$ in front of $\zeta$ is needed in order to avoid 
singularity at $\dot{\phi} = 0$ \cite{FB1,FB2}. 
Since $1 + w$ just oscillates between finite values, 
a growth of the left-hand side of (\ref{zetadef}) 
can only come from a growth of $\zeta$. 

The momentum constraint (\ref{quasimc}), can be generalized to the full three
dimensions; the resulting scalar part, separated along the lines of 
Ref. \cite{stewart}, is
\be
\dot{\Psi}_{\bf k} + \Hub \Phi_{\bf k} = -\frac{\kappa^2}{2} \left(
\dot{\phi} \delta\phi_{\bf k} + \frac{1}{k^2}
\int d^3 p (\k\cdot \p) \dot{\chi}_{{\bf p}+{\bf k}} \chi^*_{\bf p}  \right)
\equiv 
-\frac{\kappa^2}{2} \dot{\phi} \delta\phi_{\bf k} - \Z_{\chi\,{\bf k}} \; .
\label{mc3d}
\ee
Using (\ref{mc3d}), we can express $\zeta$ as
\be
(1+w) \zeta_{\bf k} = - \frac{\kappa^2}{3 \Hub} \dot \phi \, Q_{\phi\,{\bf k}} 
- \frac{2}{3\Hub} \Z_{\chi\,{\bf k}} \, .
\label{zetaquasi}
\ee
At small $k$, we estimate the integral in (\ref{mc3d}), in case (ii), as
\be
\int d^3 p (\k\cdot \p) \dot{\chi}_{{\bf p}+{\bf k}} \chi^*_{\bf p} 
\sim {\cal O} (k^2) \; ,
\label{chiinzeta}
\ee
so that the direct contribution of $\chi$ in this case is independent
of $k$ in the small-$k$ limit. 
The same conclusion was reached previously in Ref. \cite{LLMW},
albeit for a different model and by a different method. 
However, comparing (\ref{zetaquasi}) 
and (\ref{pwsQ}), we see that at $g^2 \gg \lambda$ the contribution 
from $Q_{\phi}$, which was not considered
in Ref. \cite{LLMW}, is parametrically larger than the direct contribution
of $\chi$.

Even that larger contribution is still quite small. 
Using (\ref{pwsQ2}), we obtain
\be
P_{\zeta, {\rm NG}}(k) \sim \frac{\kappa^4 \dot{\phi}^2 \phi^2}
{\Hub^2 k_*^3} \sim \frac{\phi^2}{M_{\rm Pl}^2 k_*^3} \; ,
\ee
so that the perturbation of $\zeta$ on scales larger than the Hubble scale
is
\be
\delta_H \zeta \sim \frac{\phi}{M_{\rm Pl}} \left( \frac{\Hub}{k_*} \right)^{3/2}
\sim  10^{-6} \; ,
\ee
where we have used $\phi/M_{\rm Pl} \sim 3\times 10^{-3}$ 
and $\Hub/k_* \sim 6 \times 10^{-3}$.
We doubt that such a small perturbation can cause a gravitational
instability, such as one observed numerically in Ref. \cite{FK}. Indeed,
in the following sections we will identify a much stronger infrared  effect, due
to anisotropic stress.

The form (\ref{green1}) of the Green function leaves open the possibility
of $\delta\phi$ cascading towards the infrared. Indeed, if the integral
in the argument of sine increases with time, the region of $k$ where
the regular small-$k$ limit applies will be pushed further and further
in the infrared. However, the corresponding growth of the spatial scale
of the perturbations can only occur at the relevant excitation speed, such
as the speed of sound, and so cannot overtake the cosmological horizon
(which recedes at the speed of light).

\section{Initialization of a radial field}
It is clear from Eq. (\ref{quasiN}) that to understand the effect of the
anisotropic stress, we need to understand the spatial profile of the field
$\chi$. It is useful to look first at the initial state,
because the profile of $\chi$ remains qualitatively the same throughout 
the quasilinear stage of the evolution.

A spherically-symmetric (also called radial) field $\chi$ is 
initialized according to the formula\footnote{
In numerical work, after the field $\chi$ has been initialized according to
(\ref{rchi}) we additionally subtract away its homogeneous component. This
does not matter for the present argument.}
\be
r\chi(r) = A \sum_k s_k k P_{\chi}^{1/2}(k) \sin kr  \; ,
\label{rchi}
\ee
where $k$ are the lattice momenta, quantized in units of $\pi/R$, $s_k$ are
Gaussian random numbers with zero average and unit dispersion, 
and $P_{\chi}$ is the power spectrum. The constant coefficient $A$ is to be 
chosen upon
deciding what part of the full three-dimensional (3d) variance the radial 
field will account for. The latter question arises because in the full 3d case
there surely will be other, nonradial, modes, which will carry a part (in fact,
most) of the variance.

Consider the average of the field defined in (\ref{rchi}) over the distribution
of the random numbers $s_k$:
\be
\langle \chi^2 \rangle(r) = \frac{A^2}{r^2} \frac{R}{\pi}
\int_{k_{\min}}^{k_{\max}}  dk k^2 P_{\chi}(k) \sin^2 kr  \; ,
\label{avechi2}
\ee
where we have replaced the sum with an integral, and $k_{\min}$ and $k_{\max}$
are the minimal and maximal lattice momenta.
For a field that was heavy during inflation, we use $P_{\chi}(k)$ corresponding
to the adiabatic vacuum, with an additional ultraviolet cutoff:
\be
P_{\chi}(k) = (2\pi)^{-3} (2\omega_k)^{-1} \exp(-k^2 / k_0^2) \; .
\label{Pchi}
\ee
where $\omega_k$ is the instantaneous frequency at $t = 0$. For this power
spectrum, the average (\ref{avechi2}) has the asymptotic form
\be
\langle \chi^2 \rangle(r) \approx \frac{A^2}{r^2} \frac{R}{2\pi}
\int   P_{\chi}(k) k^2 dk
\label{r2}
\ee
at large $r$. On the other hand, the full 3d variance is independent of the
coordinates and equals
\be
\langle \chi^2 \rangle_{\rm full} = 4\pi \int P_{\chi}(k) k^2 dk \; .
\label{full}
\ee

The field $\chi$ is smooth, almost constant, over a patch of radius 
$r\ll \pi/k_0$, where
$k_0$ is the ultraviolet cutoff from (\ref{Pchi}). A smooth field is
predominantly an $s$-wave, so for such $r$
we expect the average (\ref{avechi2}) to approximately equal the full
3d variance (\ref{full}). In a sense, the scale
\be
\ell_0 = \pi / k_0
\label{ell}
\ee
can be viewed as the initial correlation length of the field.
The asymptotic expression (\ref{r2}) applies at $r\gg \ell_0$, but as
$r$ approaches $\ell_0$, we expect $\langle \chi^2 \rangle(r)$ to approach
$\langle \chi^2 \rangle_{\rm full}$. Accordingly, we adopt the following
estimate for the magnitude of $\langle \chi^2 \rangle(r)$ at $r\gg \ell_0$:
\be
\langle \chi^2 \rangle(r) = \langle \chi^2 \rangle_{\rm full} \ell_0^2 /r^2 \; .
\label{est}
\ee
This corresponds to 
\be
A^2 = 8\pi^2 \ell_0^2/ R \; .
\label{A2}
\ee
A number of order one could be inserted here, but it has practically no effect
on the final results. The important point is that, at a fixed $r$, 
Eq. (\ref{est}) gives the size of fluctuations that is independent of the
spatial volume. Equivalently, the volume average
\be
\frac{4\pi}{V} \int_0^R r^2 dr \langle \chi^2 \rangle(r) =
\frac{4\pi\ell_0^2 R}{V}  \langle \chi^2 \rangle_{\rm full},
\label{chivar}
\ee
decreases with $R$ as $1/R^2$.

The $1/r^2$ dependence of the variance of a radial field can be simply
interpreted as follows. For an observer at the origin, a smooth patch of
size $\ell_0$ at distance $r \gg \ell_0$ has angular size $\ell_0 / r$.
Thus, expansion of a 3d field with correlation length $\ell_0$ will require of
order $(r/\ell_0)^2$ different spherical harmonics, with amplitudes that are
roughly of the same order.
The variance in each of these partial waves, specifically
the $s$-wave, will then account for $(\ell_0/r)^2$ of the full 3d variance.
Similar considerations apply to perturbations of the metric.

In Ref. \cite{FK}, in preparation of the initial state,
we in effect assumed that the volume averages of
$\langle \chi^2 \rangle$ and of the variances of the metric functions,
e.g. the left-hand side of (\ref{chivar}), equaled the
full 3d variances. As we now see, that was an overestimate of the size of
initial fluctuations. However, as typical in problems where
amplification of fluctuations is cut off by nonlinear effects, the initial
size is relatively unimportant: repeating the run of Ref. \cite{FK} with
initial data
of the correct size, we find that the strong gravitational effects,
which we interpreted as formation of a black hole, are merely postponed 
from $t\approx 74$ to $t\sim 105$ (in the units of Ref. \cite{FK}).

\section{Effect of the anisotropic stress}

We are now in position to estimate metric perturbations caused by the
anisotropic stress. Consider Eq. (\ref{quasiN}) of the quasilinear 
approximation. As we have seen in the previous section,
the initial profile of $\chi$ is a superposition of spherical waves and has
a $1/r$ envelope. This form is preserved during the quasilinear stage for
as long as Eq. (\ref{quasichi}) applies.
However, the correlation length of $\chi$ is now determined not by
the ultraviolet cutoff, but rather by the typical resonant momentum,
$k_*$:
\be
\ell = \pi / k_* \; .
\label{ell1}
\ee
Accordingly, instead of (\ref{est}) we now have, at $r \gg \ell$,
\be
\langle \chi^2 \rangle (r,t) = \langle \chi^2 \rangle_{\rm full} (t)
\ell^2(t) / r^2 \; .
\label{est1}
\ee
Averaging Eq. (\ref{quasiN}) over the distribution of the random initial 
data and using Eq. (\ref{est1}), we obtain
\be
r \left[ \frac{(\langle \Phi \rangle - \langle \Psi\rangle )'}{r} \right]' = 
c \kappa^2 \frac{\langle \chi^2 \rangle_{\rm full}}{r^2} \; ,
\label{aver}
\ee
where $c>0$ is a weakly time-dependent constant of order one. 
Two powers of $\ell$ in Eq. (\ref{est1}) eat away two spatial derivatives
in Eq. (\ref{quasiN}).

Eq. (\ref{aver}) is easily integrated. Using the boundary conditions
$(\Phi - \Psi)(R) = 0$, $(\Phi - \Psi)'(R) = 0$, we obtain\footnote{
Using, instead of $(\Phi - \Psi)(R) = 0$, the condition of vanishing
volume average of $\Phi - \Psi$, as in Sect. \ref{sect:quasi}, will only
change (\ref{Phi-Psi}) by an unimportant constant.}
\be
\aPhi - \aPsi = {1\over 2} 
c \kappa^2 \langle \chi^2 \rangle_{\rm full} 
\left\{ \ln\frac{R}{r} + {1\over 2} \frac{r^2 - R^2}{R^2} \right\} \; .
\label{Phi-Psi}
\ee
Thus, the average gravitational potentials are more long-ranged than
the average of $\chi^2$ itself, Eq. (\ref{est1}).

Combining Eq. (\ref{Phi-Psi}) with the momentum constraint (\ref{quasimc}), we
can obtain estimates for $\aPhi$ and $\aPsi$ separately, at
least for $r > \ell$. We will need to integrate Eq. (\ref{quasimc}) so as 
to remove the spatial derivative. The second term on the right hand side can 
only lead to a short-range, i.e. $1/r^2$, contribution, while the first term 
leads to a contribution
that is either short-range or suppressed by several powers of $M_{\rm Pl}$.
So, for our present purposes we can take
\be
\langle {\dot \Psi} \rangle + \Hub \aPhi \approx 0 \; .
\label{mc1}
\ee
If, as a result of parametric resonance, the field $\chi$ grows with
a characteristic exponent $\mu$, the potentials $\Phi$ and $\Psi$, as 
(\ref{Phi-Psi}) and (\ref{mc1}) show,
grow with characteristic exponent $2\mu$. Then, from
(\ref{mc1}) we see that
\be
\aPsi \sim \frac{\Hub}{\mu m_0} \aPhi \ll \aPhi \; ,
\label{PsillPhi}
\ee
where $m_0\sim \sqrt{\lambda}\phi(0)$ is the frequency (in conformal time) of the 
inflaton's oscillations. Therefore,
\be
\aPhi \sim   \frac{\langle\chi^2 \rangle_{\rm full}}{ M_{\rm Pl}^{2} }
\ln\frac{R}{\ell} \; ,
\label{Phi}
\ee
and then (\ref{PsillPhi}) is an estimate for $\aPsi$. 
We see that, up to a weak logarithmic factor, the average potentials are 
independent of $R$. We stress, however, that
these estimates apply only in quasilinear theory.

The approximate
volume independence of $\aPhi$ and $\aPsi$ during the resonant growth of $\chi$
is confirmed by numerical integrations. Integrations were done using the
algorithm described in Ref. \cite{FK} and are fully nonlinear. 
For example, Fig. \ref{fig:psi} shows $\delta_H\psi / \psi(R)$ 
for $g^2/\lambda = 110$ and
$\lambda' = 0$. Here $\psi(R)$ is the conformal factor at the outer end
of the grid, and $\delta_H\psi$ is defined by
\be
(\delta_H\psi)^2 = 4\pi \int_{k_{\min}}^{\Hub} dk k^2 P_{\psi}(k) \; ,
\label{svar}
\ee
where $P_{\psi}(k)$ is the power spectrum of $\psi -  \psi(R)$, and
$k_{\min}$ is the minimal lattice momentum.
Fourier components of $\psi -  \psi(R)$, required for calculation of the
power spectrum, are determined via a sine transform.
During the resonant growth of $\chi$, the metric functions $\Phi$ and
$\Psi$ rapidly become smooth and can therefore be regarded as self-averaging.
For this reason, the quantity plotted in Fig. \ref{fig:psi} should give a good
idea of the typical values of $\aPsi$.

Numerical values of $\aPhi$ and $\aPsi$ predicted by the estimates 
(\ref{Phi}) and (\ref{PsillPhi}) are quite small. The usual estimate 
\cite{KT2,KTGRAV} for the maximum value of
$\langle \chi^2 \rangle_{\rm full}$ is
\be
\max\langle \chi^2 \rangle_{\rm full} 
\sim \frac{\lambda \phi^2}{g^2}
\sim \frac{\lambda}{g^2} 
\times \frac{10^{-2} M_{\rm Pl}^2}{\ln^2\lambda}  \; ,
\label{maxchi2}
\ee
so for $\lambda=10^{-13}$ and $g^2/\lambda \sim 100$ we obtain
\be
\aPsi \ll \aPhi \sim 10^{-7} \ln (R/\ell) \; .
\label{PhiandPsi}
\ee
From Fig. \ref{fig:psi}, we see that the lattice result by far exceeds
the estimate (\ref{PhiandPsi}), even during the stage of exponential
growth, which one might want to identify with the quasilinear stage.
The reason is that, because of the $1/r^2$ form of $\langle \chi^2 \rangle$,
the quasilinear approximation, strictly speaking, breaks down quite early
on---when $\langle \chi^2 \rangle$ becomes of order (\ref{maxchi2}) at
$r\sim \ell$. The growth of various quantities can still continue, however,
because in most of the space $\langle \chi^2 \rangle$ is still much smaller
than that. (We discuss this point further in the next section.)

During the entire exponential growth, the two curves corresponding
to different box
sizes coincide with excellent precision, in agreement with the approximate
volume independence of $\aPsi$ derived analytically. The time when the curves
part corresponds to breakdown of 
the quasilinear theory in most of the space. At that time, volume dependence 
of the results is not surprising: nonlinear gravitational phenomena, such
as gravitational collapse,
depend on the total amount of matter available and will be more pronounced in 
a larger volume.

\section{Further nonlinear effects} \label{sect:nonl}

The driving force due to local scattering in Eq. (\ref{quasiphi}) 
can be estimated as
\be
g^2 \phi \langle \chi^2 \rangle \sim g^2 \phi \ell^2 
 \langle \chi^2 \rangle_{\rm full} / r^2 \; .
\label{local}
\ee
Fluctuation of $\phi$ caused by this driving force at 
$r\sim \ell$ is
\be
\delta \phi \sim 
\frac{g^2 \langle \chi^2 \rangle}{\lambda \phi} \; .
\ee
This becomes of order $\phi$ when $\langle \chi^2 \rangle$ reaches
the value (\ref{maxchi2}).
After that $\langle \chi^2 \rangle$ does not grow any further
at $r\sim \ell$ but continues to grow at $r\gg \ell$.

From our lattice data, it appears that this later growth is a combination
of the usual resonant growth and rescattering of fluctuations from smaller
$r$, where they are already large.. As a result,
the region where $\langle \chi^2 \rangle$ is large propagates
to larger $r$, so that the quasilinear form (\ref{est1}) no longer applies. 
From Eq. (\ref{quasiN}), we see that the growth of this region
leads to a further growth of metric perturbations.
If the total volume is large enough, these can eventually become
sufficiently large to cause a gravitational instability.

The time when the resonant growth of $\chi$ has ended in most of the space
corresponds to a short plateau around $t=250$ in Fig. \ref{fig:psi}.
The subsequent growth
of $\delta_H\psi$ has a different origin: we interpret it is a result of
an additional production of $\delta\phi$ by another, weaker resonance, due 
to the $\lambda\phi^4/4$ interaction. During
this further growth, we observe formation of a black hole---the lapse
$N$, as a function of $r$, crosses zero.\footnote{
One may, and some physicists did, debate whether this is an adequate
definition of a black hole. Our main argument remains the same as in Ref.
\cite{FK}: for the metric (\ref{metric}), a sphere on which $N=0$ cannot
be crossed by a classical particle, and is therefore a horizon. On the
other hand,
quantum particles, or classical waves, can transport energy across the 
horizon (similarly to Hawking radiation); we suppose this transport is
the cause of growth or evaporation of black holes in our simulations.}

The presence of the second stage of growth in Fig. \ref{fig:psi} leads
to an important conclusion. If some decay channel
becomes blocked while the inflaton's homogeneous component still has not
decayed completely, the inflaton will ``wait'' until a new channel
opens up. At least in our spherically-symmetric geometry, and in the model
considered here, its decay via the new channel still causes significant metric 
perturbations. This conclusion holds also in the case when the decay
of the inflaton into $\chi$ is blocked by a $\lambda'\chi^4/4$
self-interaction (cf. Ref. \cite{PR}): we show the result 
for $g^2/\lambda = 8$ (a stronger resonance)
and $\lambda' /\lambda = 1000$ by short dashes in Fig. \ref{fig:psi}.
In this case, the decay into $\chi$ is cut off at much smaller values
of metric perturbations, and the interpretation of the second stage
of growth as the $\lambda\phi^4$ resonance is unambiguous.

\section{Conclusion} \label{sect:conc}
In this paper we have studied the influence of nonlinear effects, arising
during decay of the inflaton, on metric perturbations.
Although we restricted our attention to a quartic potential for the inflaton,
results for other models of chaotic inflation can be obtained by similar methods, 
and we believe they will be similar in nature.

Early stages of the inflaton decay can be studied analytically, using
the quasilinear approximation, at least under the restriction to spherical
symmetry (a similar approximation may exist also in
the full 3d case). The quasilinear approximation retains quadratic terms 
in the decay products; these are the leading terms affecting
metric perturbations after inflation
in the case when fields interacting with the inflaton do not have 
sizeable homogeneous components.

Using the quasilinear approximation, we have identified two effects
that could potentially produce large-scale metric perturbations after 
the end of inflation. We have found that the first of these 
effects---scattering of the second field $\chi$ off the inflaton's homogeneous 
component---does
not cause significant super-Hubble perturbations in the quasilinear regime.
However,
the second type of effect---metric perturbations caused by the anisotropic
stress---has been found to be important.

It is interesting to contemplate how this latter effect is able to avoid
the restriction on super-Hubble physics imposed by causality.
To this end, note that a spherical wave has a large degree of {\em angular} 
coherence: 
even for a wave with momentum much larger than the Hubble parameter,
angular coherence extends over scales that are not in causal contact. 
Parametric amplification of such fluctuations
is not restricted by causality (just as it is not restricted during
inflation). 

Persistence of angular coherence of fluctuations in the full three 
dimensions corresponds to a large mean-free path. As we have discussed in
Sect. \ref{sect:sphe}, 
this is indeed likely to be the case at the early stages of 
reheating in weakly-coupled models. However, because restriction to
spherical symmetry explicitly breaks translational invariance, it is not
immediately obvious how our results can be transplanted to the full 3d case.
Indeed, the $r$ dependence of the ensemble averages $\aPhi$ and
$\aPsi$ (cf. (\ref{Phi-Psi})) is a direct consequence of the translational
non-invariance and is not expected to persist in three dimensions. 
Instead, the 3d picture
that we have in mind is a sort of ``metric turbulence'', in which a (nearly)
spherically-symmetric fluctuation in a sphere of radius $R_1$ sits on top
of a fluctuation in a sphere of a larger radius $R_2$. Then, instead of
$\aPhi$ and $\aPsi$ we should consider the total perturbation from all scales
shorter than some $R$, e.g.
\be
\delta_R \Phi = \langle [\Phi(\x) - \Phi(\x + \R)]^2 \rangle^{1/2} \; ,
\label{deltaR}
\ee
so for example the quasilinear result (\ref{Phi}) will correspond to 
\be
\delta_R \Phi \sim \frac{\langle\chi^2 \rangle_{\rm full}}{ M_{\rm Pl}^{2} }
\ln\frac{R}{\ell} 
\label{deltaPhi}
\ee
(with an approximately scale-invariant 3d spectrum
$P_{\Phi}(k) \propto k^{-3} \ln k$).

Perturbation (\ref{deltaPhi}) is by itself quite small.
We have seen, however, that in
spherical symmetry a much larger perturbation is possible, due to 
spreading out of the region where fluctuations of $\chi$ are large.
In Sect. \ref{sect:nonl}, we have argued that this is
indeed the mechanism of gravitational instability and formation of black holes
observed in our lattice studies. 
Because this mechanism involves already large field fluctuations, we cannot
exclude the possibility that in the full three dimensions scattering of these 
large fluctuations
destroys angular coherence, by breaking the $s$-wave
into higher partial waves, and the formation of black holes is prevented.
However, if black holes do form in the full 3d by the mechanism identified here,
the question remains what is the spectrum of metric perturbations after
these black holes evaporate.

This work was supported 
in part by the U.S. Department of Energy through Grant DE-FG02-91ER40681 
(Task B).

\end{document}